# Visualizing an adjustable $WO_3$/p-GaN heterojunction

Yuliang Chen, Changlong Hu, Guobin Zhang, Chongwen Zou[*]

*National Synchrotron Radiation Laboratory, University of Science and Technology of China, Hefei, 230029, China*

Corresponding Authors: czou@ustc.edu.cn

**Abstract:**

**The p-n junctions based on typical semiconductors are the elementary units for the modern electronic devices and chip industry. While the rectification property of those p-n junction is usually fixed once the unit is fabricated. Here, we proposed an adjustable n-$WO_3$/p-GaN heterojunction with controllable electronic properties. For the prepared n-$WO_3$/p-GaN heterojunction, it is almost transparent and shows typical p-n junction rectification. While if gradually doping some hydrogen atoms into $WO_3$ layer by a facile electron-proton synergistic route, the heterojunction can be turned dynamically from the typical p-n junction (n-$WO_3$/p-GaN) to standard Schottky contact ($H_xWO_3$/p-GaN) step by step. More importantly, this evolution can be directly visualized by eyesight due to the pronounced electrochromic characteristic of $WO_3$ layer. By connecting two $H_xWO_3$/p-GaN heterojunctions, the controllable bi-functional rectification can be achieved. In addition, the $H_xWO_3$/p-GaN heterojunction can recovered to the original p-n jucntion just by annealing at ambient, demonstrating the heterojunction is controllable and reusable. The current study will open up tremendous opportunities for dynamic electronic devices in the future.**

**Introduction**

Heterojunctions, the interface between different materials, where the translational symmetry is broken, always create many exotic phenomena. One of the famous examples is that the two-dimensional electron gas (2DES) appearing at the interface of heterojunction shows the quantum Hall effect, such as GaAs/AlGaAs, HgTe/CdTe.[1, 2] As a popular field, the topological insulator, whose surface metallic state can be considered as the interface of vacuum and the bulk material, is still heterojunction intrinsically, such as $Bi_2Se_3$, $Bi_{1-x}Sb_x$.[3, 4] Moreover, p-n junctions, the fundamental units for logical circuits, have been revolutionizing our present information era.

The typical property of p-n junction is the so-called rectifying effect, which means electrons can only go through the n-type terminal to p-type terminal under a forward bias voltage. However, the reverse voltage would produce limited current. To form p-n junctions, the doping technology is the spirit, which directly determines the property or performance of the rectifying result. For example, in the traditional silicon-based semiconductor industry, doping donor impurity (eg., N, P) in pristine Si crystal is to get n-type Si. While in contrast, acceptor impurity (eg., B) is doped for getting p-type Si. This doping technology for Si crystal is always associated with extreme conditions such as high vacuum, high energy, or high temperature for ion implantations. Subject to these high energy-consumption processes, the produced p-n junctions are static, which means that once the doping is selected and the p-n junction is fabricated, the p-n junction would have the fixed rectification property and no feasible way can adjust its property on purpose.

The recently proposed electron-proton synergistic effect conducting in separated electron and proton sources supplies a low-energy scale environment, which has been verified an effective electron doping for some oxide materials in ambient conditions.[5, 6] Moreover, the influence of synergistic effect can be erased by suitable temperature annealing, making the facile electron doping process reversible and controllable. Due to the pronounced electrochromic property of $WO_3$ layer, hydrogen doping for $WO_3$ layer can be directly observed by eyesight and induce the variable electric properties at the same time.[7]

Here, we proposed a dynamic electron doping for $WO_3$ layer and observed the rectification change of p-n junction at $WO_3$/p-GaN interface. Experimentally, a $WO_3$ film was prepared on p-GaN substrate by reactive magnetron sputtering to form a typical p-n junction with the rectification ratio of more than 100. While if doping H atoms into $WO_3$ layer by the electron-proton synergistic strategy in acid solution, a phase transition was observed in $WO_3$ film accompanying the clear color change from transparency to blue. Resultantly, the p-n contact ($WO_3$/p-GaN) was also transferred to Schottky contact ($H_xWO_3$/p-GaN) gradually. It was noticed that the modulated heterojunctions can be restored by high-temperature annealing in air, making the junction device reusable. This dynamically reusable $WO_3$/p-GaN heterojunction is promising for functional electronic devices in the future.

**Experimental and results**

A thin $WO_3$ film was deposited by reactive magnetron sputtering on a p-type Mg-doped GaN substrate, as shown in Fig 1a. The work function ($W_F$) for GaN and $WO_3$ is 7.5 eV and 6.59 eV, and the band gaps ($E_g$) are 3.4 eV and 2.9 eV, respectively. These values are listed in the band alignments as shown in Fig 1b, which hints the interface of p-GaN and $WO_3$ could be a p-n junction. Indeed, the measured I-V curve in Fig 1c shows a good rectifying effect with the ratio of more than 100 as shown in Fig 1d.

Our previous report has proved that the electron-proton synergistic effect can realize a MIT in $WO_3$.[7] In Fig 2a, when the $WO_3$ film contacted with a Zinc particle (yellow square) is immersed into an acid solution, the electrons supplied by Zn and the free protons in an acid solution can be doped into $WO_3$ film together, resulting in the formation of $H_xWO_3$. Though the whole synergistic processes happened rapidly, they can be controlled by changing the contact time carefully between Zn particle and $WO_3$ film, which means the doped concentration can be adjusted on purpose. In Fig 2b, from bottom to top, the color changing from colorless to dark blue indicates the increase of doped concentration.

At differently doped concentrations, we tested the I-V curves of the heterojunction as shown in Fig 2c. A gradual transition from p-n contact to Schottky contact with increasing concentration was observed. The reason should be attributed to the MIT of $WO_3$ layer due to the hydrogenation treatment by the electron-proton synergistic doping. More importantly, it should be emphasized that the doped $H_xWO_3$ can be restored completely by annealing as shown in Fig 2d. Therefore, this heterojunction can be processed repeatedly. On the other hand, to get different doped concentration, besides controlling the contact time, controlling the annealed time is also an alternative way.

Furthermore, if we combine multiple $WO_3$/p-GaN heterojunctions, more functions can be achieved. For example, a controllable rectification at both forward and reverse biases can be realized as shown in Fig 2e. The $WO_3$ films were deposited at eight isolated areas on the same p-GaN substrate and were doped at different concentrations by electron-proton synergistic effect, showing different colors (the inset of Fig 2e). The top four areas are labeled by numbers "0-3", and labeled by alphabets "a-d" for the bottom ones. Forward (reverse) bias is applied to the top (bottom) films. In Fig 2e, it can be observed that the current not only can be depressed at both forward and reverse bias voltages (curve "3-d") but also can be active with the bi-functional way at the same time (curve "0-a"). The typical rectification p-n junction was also developed, even can realize different degrees of rectification (curves "0-d", "1-d", "2-d"). These results shown in Fig 2e indicate the rectification performance of $WO_3$/p-GaN heterojunction has high flexibility upon the hydrogenation of $WO_3$ layer by the facile electron-proton synergistic doping route in ambiance.

To conclude, we have fabricated a $WO_3$/p-GaN heterojunction, which shows the typical p-n rectification performance. By using the electron-proton synergistic doping route, the $WO_3$ layer can be dynamically hydrogenated, which can be observed by eyesight due to the pronounced electrochromic behavior of WO3 layer upon the electron doping. This dynamic hydrogenation results in the evolution from typical p-n contact ($WO_3$/p-GaN) to Schottky contact ($H_xWO_3$/p-GaN). Besides, more functional properties can be realized if combining multiple $H_xWO_3$/p-GaN heterojunctions. More importantly, the H-doped $WO_3$ layer can be restored by high-temperature annealing in air, which makes the $H_xWO_3$/p-GaN heterojunction can be transferred to the original $WO_3$/p-GaN. This controllable modulation of $WO_3$/p-GaN heterojunction demonstrates the giant potential for functional electronic devices.

**Characterizations**

WO$_3$ thin films were prepared on p-type Mg-doped GaN single-crystal substrates by reactive magnetron sputtering in an argon–oxygen atmosphere at 200℃ with a stoichiometry WO$_3$ target. Before loading the p-GaN substrates into the chamber, they were ultrasonically cleaned with acetone and ethanol and then rinsed several times with de-ionized water. The flow rates of argon and oxygen were fixed at 6.0 and 1.0 SCCM, respectively. The radio-frequency sputtering power during the deposition was maintained at 80 W. Under these conditions, the prepared WO$_3$ thin films were amorphous. The amorphous structure is in favor of the ions/protons insertion into WO$_3$ film, also triggering the quick electrochromic results. By covering a silicon stencil on the top of a p-GaN substrate, the eight WO$_3$ films were deposited on the same substrate. The purity of Zinc particle is >99.99%. The 2%wt H$_2$SO$_4$ solutions were adopted in all experiments The I-V properties were conducted on Keithley 2450 sourcemeters.

**References**

1. Tsui DC, Stormer HL, Gossard AC. Two-Dimensional Magnetotransport in the Extreme Quantum Limit. *Physical Review Letters* 1982, **48**(22)**:** 1559-1562.

2. Konig M, Wiedmann S, Brune C, Roth A, Buhmann H, Molenkamp LW*, et al.* Quantum spin hall insulator state in HgTe quantum wells. *Science* 2007, **318**(5851)**:** 766-770.

3. Zhang H, Liu C-X, Qi X-L, Dai X, Fang Z, Zhang S-C. Topological insulators in Bi2Se3, Bi2Te3 and Sb2Te3 with a single Dirac cone on the surface. *Nat Phys* 2009, **5**(6)**:** 438-442.

4. Hsieh D, Qian D, Wray L, Xia Y, Hor YS, Cava RJ*, et al.* A topological Dirac insulator in a quantum spin Hall phase. *Nature* 2008, **452**(7190)**:** 970-974.

5. Chen Y, Wang Z, Chen S, Ren H, Wang L, Zhang G*, et al.* Non-catalytic hydrogenation of VO2 in acid solution. *Nature communications* 2018, **9**(1)**:** 818.

6. Xie L, Zhu Q, Zhang G, Ye K, Zou C, Prezhdo OV, *et al.* Tunable Hydrogen Doping of Metal Oxide Semiconductors with Acid-Metal Treatment at Ambient Conditions. *Journal of the American Chemical Society* 2020, **142**(9)**:** 4136-4140.

7. https://arxiv.org/abs/2005.08819

**Acknowledgments**

This work was partially supported by the National Key Research and Development Program of China (2016YFA0401004), the National Natural Science Foundation of China (11574279, 11704362), the funding supported by the Youth Innovation Promotion Association CAS, the Major/Innovative Program of Development Foundation of Hefei Center for Physical Science and Technology. This work was partially carried out at the USTC Center for Micro and Nanoscale Research and Fabrication. The authors also acknowledged the supports from the Anhui Laboratory of Advanced Photon Science and Technology. The approved beamtime on the XMCD beamline (BL12B) in National Synchrotron Radiation Laboratory (NSRL) of Hefei was also appreciated.

**Competing financial interests**

The authors declare no competing financial interests.

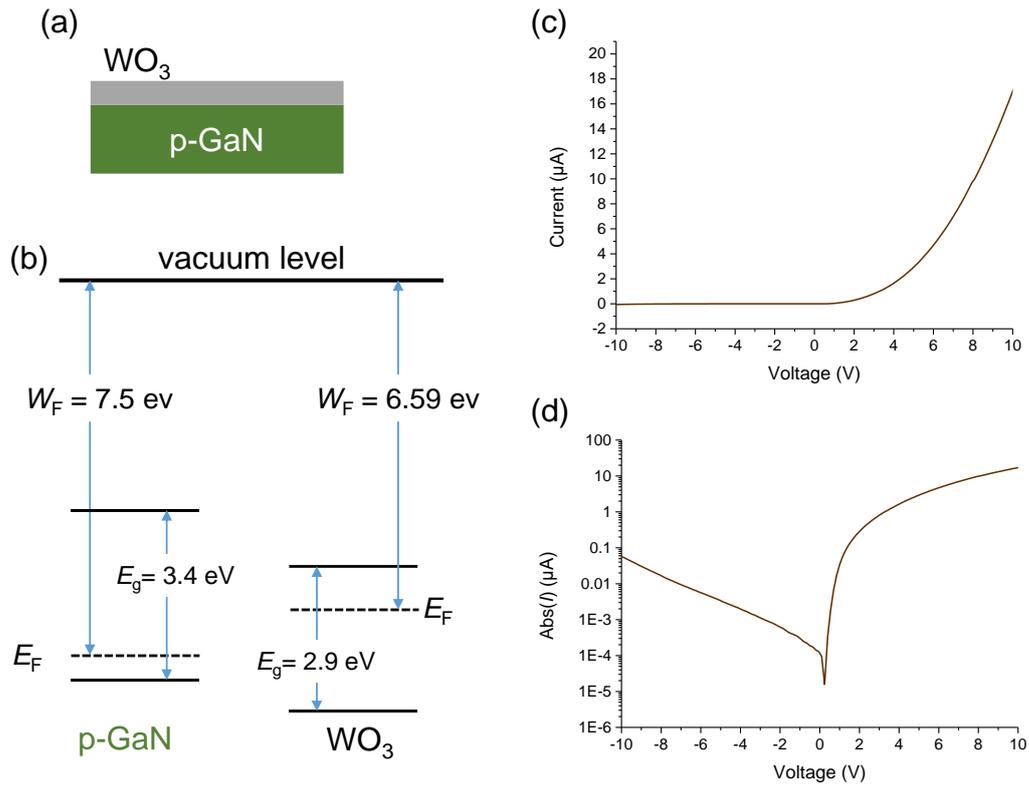

Figure 1. (a) WO$_3$ film was deposited on a p-type GaN substrate by reactive magnetron sputtering. (b) The band alignments for p-GaN and WO$_3$. (c-d) The I-V curve of WO$_3$/p-GaN heterojunction in linear scale (c) and logarithmic scale (b), respectively.

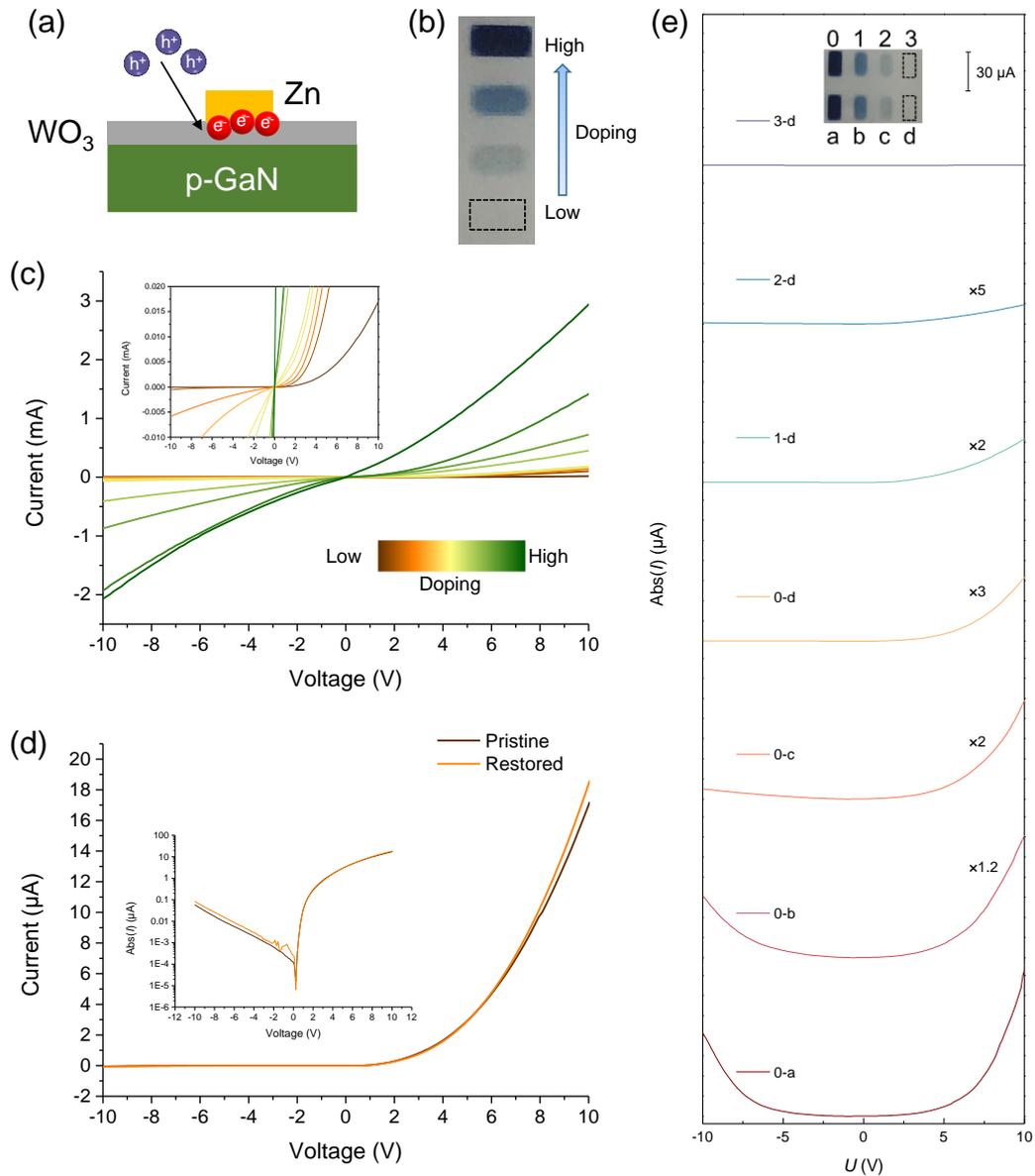

Figure 2. (a) The scheme for the electron-proton synergistic doping route. The whole system is immersed in an acid solution. The purple bubbles are free protons in the acid solution and the yellow column is a Zinc particle for supplying electrons. (b) The electron-proton synergistic doping can be observed by eyesight due to the electrochromic effect of $WO_3$ layer. The size of the single doped area is $4 \times 2$ mm$^2$. (c) The I-V curves of the heterojunction with respect to different doped concentration. The details are in the inset. (d) The pristine rectification of p-n junction is reversible by annealing in air. The inset is plotted in logarithmic scale. (e) The I-V curves between different films. The certain curves were multiplied by a corresponding coefficient for notice. Inset: the eight $WO_3$ films were deposited on the same p-GaN substrate and were doped at different concentrations, which were labeled by numbers and alphabets respectively.